\documentclass{article}

\usepackage[dvipsnames]{xcolor}
\usepackage{graphicx}
\usepackage{booktabs}
\usepackage{subcaption}
\usepackage{amsmath}
\usepackage{amssymb}
\usepackage{bm}

\newcommand{\archname}{Serpent}
\newcommand{\archb}{\archname{}-B}
\newcommand{\archl}{\archname{}-L}
\newcommand{\archh}{\archname{}-H}
\newcommand{\block}{S-block}
\newcommand{\blockb}{\block{} B}
\newcommand{\blockd}{\block{} D}
\newcommand{\blocku}{\block{} U}

\usepackage{arxiv}
\usepackage{graphicx, subcaption}
\usepackage[normalem]{ulem}
\usepackage[utf8]{inputenc} 
\usepackage[T1]{fontenc}    
\usepackage[hyperfootnotes=false]{hyperref}       
\usepackage{url}            
\usepackage{booktabs}       
\usepackage{amsfonts}       
\usepackage{nicefrac}       
\usepackage{microtype}      
\usepackage{xcolor}         
\usepackage{algpseudocode}
\usepackage{bm}
\usepackage{amssymb}  
\usepackage{amsthm}
\usepackage{amsmath}
\newcommand{\vct}[1]{\bm{#1}}
\newcommand{\mtx}[1]{\bm{#1}}

\usepackage{footmisc}
\DefineFNsymbols{mySymbols}{{\ensuremath\dagger}{\ensuremath\ddagger}\S\P
	*{**}{\ensuremath{\dagger\dagger}}{\ensuremath{\ddagger\ddagger}}}
\setfnsymbol{mySymbols}

\title{\archname{}: Scalable and Efficient Image Restoration via Multi-scale Structured State Space Models}

\author{%
	Mohammad Shahab Sepehri\\
	Dept. of Electrical and Computer Engineering\\
	University of Southern California\\
	Los Angeles, CA \\
	\texttt{sepehri@usc.edu} \\
	\And
	Zalan Fabian\\
	Dept. of Electrical and Computer Engineering\\
	University of Southern California\\
	Los Angeles, CA \\
	\texttt{zfabian@usc.edu} \\
	\And
	Mahdi Soltanolkotabi \\
	Dept. of Electrical and Computer Engineering\\
	University of Southern California\\
	Los Angeles, CA \\
	\texttt{soltanol@usc.edu}\\
	}

\begin{document}
	
	\maketitle
	\begin{abstract}
The landscape of computational building blocks of efficient image restoration architectures is dominated by a combination of convolutional processing and various attention mechanisms. However, convolutional filters, while efficient, are inherently local and therefore struggle with modeling long-range dependencies in images. In contrast, attention excels at capturing global interactions between arbitrary image regions but suffers from a quadratic cost in image dimension. In this work, we propose Serpent, an efficient architecture for high-resolution image restoration that combines recent advances in state space models (SSMs) with multi-scale signal processing in its core computational block. SSMs, originally introduced for sequence modeling, can maintain a global receptive field with a favorable linear scaling in the input size. We propose a novel hierarchical architecture inspired by traditional signal processing principles that converts the input image into a collection of sequences and processes them multi-scale. Our experimental results demonstrate that Serpent can achieve reconstruction quality on par with state-of-the-art techniques while requiring orders of magnitude less compute (up to $150$ fold reduction in FLOPS) and a factor of up to $5\times$ less GPU memory while maintaining a compact model size. The efficiency gains achieved by \archname{} are especially notable at high image resolutions. The source code and pretrained models are available at \href{https://github.com/AIF4S/Serpent}{https://github.com/AIF4S/Serpent}.
\end{abstract}
	\section{Introduction}
Image restoration is aimed at recovering a clean image from its degraded counterpart, encompassing crucial tasks such as superresolution \cite{kim2016accurate, wang2021real}, deblurring \cite{tao2018scale,zhang2020deblurring}, inpainting \cite{yu2018generative,lugmayr2022repaint} and JPEG compression artifact removal \cite{galteri2017deep}. End-to-end deep learning techniques that directly learn the mapping from corrupted images to their clean counterparts are the current state-of-the-art in most image recovery tasks. The careful design of such architectures has attracted considerable attention in recent years, and is crucial for the performance and efficiency of image restoration methods.

Architectures composed of convolutional building blocks have achieved great success in a multitude of image restoration problems \cite{mao2016image, tian2020deep} thanks to their compute efficiency. However, convolutional neural networks (CNNs) are limited in low-level vision tasks by two key weaknesses. First, convolutional filters are content-independent, that is different image regions are processed by the same filter. Second, convolutions have limited capability to model long-range dependencies due to the small size of kernels, requiring exceedingly deeper architectures to increase the receptive field. 

More recently, Transformer architectures such as the Vision Transformer \cite{dosovitskiy_image_2020}, have shown enormous potential in a variety of vision problems, including dense prediction tasks such as image restoration \cite{zamir2022restormer,wang2022uformer,liang_swinir_2021, zhao2023comprehensive}. Vision Transformers split the image into non-overlapping patches, and process the patches in an embedded token representation. Transformers lack the strong architectural bias of convolutional networks and thus have better scaling properties when trained on massive datasets. Furthermore, their core building block, self-attention, is not limited by the shortcomings of convolutions and can effectively model long-range dependencies in images. However, the added benefit of model expressiveness is overshadowed by the steep, quadratic scaling of compute with image resolution. A flurry of activity has emerged around combining the efficiency of convolutions with the representation power of Transformers \cite{zamir2022restormer, liang_swinir_2021, chen2022cross}, yielding various architectures along the efficiency-performance trade-off curve induced by these two dominant computational building blocks. For instance,  Restormer \cite{zamir2022restormer} achieves linear complexity by applying self-attention in the channel dimension, however it still requires high memory and computation.

Very recently, Mamba \cite{gu_mamba_2023}, a novel variant of structured state space models (SSMs) \cite{gu2021efficiently, gu2021combining} has been introduced for computationally efficient modeling of long sequences in natural language processing (NLP) through selective state spaces. As opposed to Transformers, the compute cost of Mamba scales linearly with sequence length, while preserving the ability to model long-range dependencies. In particular, SSMs take in a $1$-D input sequence, where each element of the array can interact with any of the previously scanned elements through a low-dimensional hidden state. Some early work in adapting SSMs with selective state spaces to visual representation learning \cite{yu_vmamba_2024,zhu_visionmamba_2024} have demonstrated promising results in model efficiency and scalability. Despite being less explored, SSMs have immense potential in dense prediction tasks such as image restoration, especially at high resolutions due to two key factors. First, high-resolution reconstruction often requires efficient modeling of long-range pixel dependencies, a capability lacking in traditional CNNs. Second, as the input dimension increases, compute and memory efficiency become essential. The quadratic scaling of standard self-attention is prohibitive even on contemporary state-of-the-art GPUs at high resolutions.

 \begin{figure}
	\centering
	\begin{subfigure}{0.42\textwidth}
		\centering
		\includegraphics[width=\textwidth]{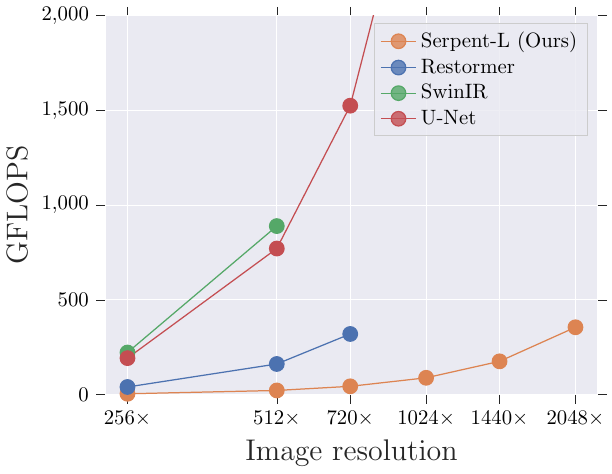}
	\end{subfigure}
	\begin{subfigure}{0.42\textwidth}
		\centering
		\includegraphics[width=\textwidth]{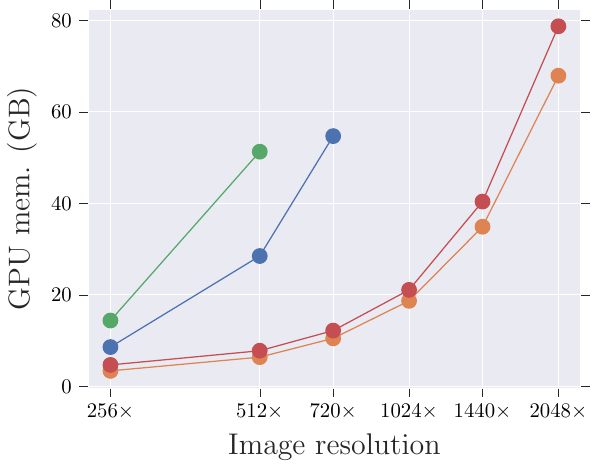}
	\end{subfigure}
	\caption{Efficiency of \archname{}: \archname{} out-scales state-of-the-art reconstruction techniques in terms of FLOPS, especially at high image resolutions (\underline{left}). Our technique matches the performance of state-of-the-art methods while utilizing $5\times$ less GPU memory during training, matching the scaling of fully convolutional architectures (\underline{right}). We plot results with batch size $1$ on a single H100 GPU with $80$ GB memory. Missing data points indicate that the given model is out of memory with the specific input resolution. \label{fig:frontpage}}
\end{figure}

In this work, we propose \archname{}, a novel architecture for efficient image restoration that leverages state space models capable of modeling intricate long-range dependencies in high-resolution images with a favorable linear scaling in input dimension. Our architecture builds upon state space models with selective scan, such as Mamba, while carefully addressing  considerations specific to dense vision tasks such as image restoration.
Our contributions are as follows:
\begin{itemize}
	 \item To the best of our knowledge, our work is among the first to successfully leverage SSMs with selective scans for high-resolution image restoration with a focus on model efficiency. Motivated by traditional signal processing principles and successful architectures in image segmentation and reconstruction, we design a hierarchical architecture that processes the input image in a multi-scale fashion. We incorporate Mamba-like blocks tailored for processing images that demonstrate favorable compute and memory scaling with respect to input size, while maintaining high reconstruction quality.
	 \item Our experiments demonstrate that \archname{} can match the image quality of state-of-the-art techniques, but with orders of magnitude lower compute (up to $150\times$ reduction in FLOPS compared to SwinIR, see Figure \ref{fig:frontpage}, left), reduced training speed ($5 \times $ faster than SwinIR \cite{liang_swinir_2021} and $3\times$ faster than Restormer \cite{zamir2022restormer}) and rapid inference (able to process $35\times$ more images per second than SwinIR). 
	 \item Our efficient design allows for low GPU memory requirements and a compact model size. In particular, \archname{} is able to reconstruct $2048 \times$ resolution images without resorting to activation checkpointing or patch-by-patch reconstruction. We achieve a reduction of $5\times$ in GPU memory compared to Restormer at $512\times$ resolution (see Figure \ref{fig:frontpage}, right), allowing for more modest hardware infrastructure when deployed on high-resolution images.
\end{itemize}
	\section{Background}
\subsection{State space models} State Space Models (SSMs) \cite{gu2021efficiently, gu2021combining} are a recent class of sequence models  with roots in classical state space models in control theory \cite{kalman1960new} that combine the characteristics of RNNs and convolutional networks. Notably, their output can be computed efficiently as either a convolution or recurrence with linear scaling in sequence length. In particular, the relationship between the input and output sequence is described through the discretization of a continuous-time linear system
\begin{align}\label{eq:lti_cont}
		\dot{\vct{x}}(t) &= \mtx{A}\vct{x}(t) + \mtx{B}u(t) \nonumber \\
		\vct{y}(t) &= \mtx{C}\vct{x}(t) + Du(t),
\end{align}
where $u \in \mathbb{R}$ is the input, $\vct{x} \in \mathbb{R}^N$ is the hidden state, $y \in \mathbb{R}$ is the output and $\mtx{A} \in \mathbb{R}^{N\times N}$, $\mtx{B} \in \mathbb{R}^{N \times 1}$, $\mtx{C} \in \mathbb{R}^{N \times 1}$, $D \in \mathbb{R}$. Effectively, the hidden state compresses  information about the past that can be accessed when processing the present input in the sequence. In discrete-time domain, these equations take the form
\begin{align*}
	\label{eq:rnn}
		\vct{x}_k &= \bar{\mtx{A}} \vct{x}_{k-1} + \bar{\mtx{B}}u_k \nonumber\\
		y_k &= \mtx{C}\vct{x}_k,
\end{align*}
where $ \bar{\mtx{A}}, \bar{\mtx{B}}$, and $ \bar{\mtx{C}}$ are the results of applying some discretization rule, such as zero-order hold, for a given time step $\Delta$ and typically $D$ is omitted. 
In what is known as \textit{CNN mode}, the output is obtained via the convolution
\begin{equation*}
	\label{eq:cnn}
		\mtx{K} = (\mtx{C}\bar{\mtx{B}}, \mtx{C}\bar{\mtx{A}}^1\bar{\mtx{B}}, \dots, \mtx{C}\bar{\mtx{A}}^L\bar{\mtx{B}}),~~
		\vct{y} = \vct{u} * \mtx{K},
\end{equation*}
where $L$ is the length of the input sequence $\vct{u}$. In CNN mode, SSMs scale linearly in sequence length making them much more efficient than Transformers, which are quadratic in input length. Moreover, unlike Transformers, SSMs do not need to calculate quadratic attention matrices, and as a results they are more memory efficient.
Structured SSMs (S4) \cite{gu2022parameterization, gupta2022diagonal, smith2022simplified} impose a specific structure on the system in Eq. \eqref{eq:lti_cont}, most commonly working under the assumption that $\mtx{A}$ is diagonal. Therefore $\mtx{A}, \mtx{B}, \mtx{C}$ can each be represented in $N$ dimensions.

\subsection{Selective SSMs} SSMs are represented by linear time-invariant (LTI) systems with constant dynamics unable to affect the hidden states in an input-dependent way. Mamba \cite{gu_mamba_2023} adds context-awareness to SSMs via selectivity: $\mtx{B}$, $\mtx{C}$ and $\Delta$ are replaced by input-dependent mappings $s_{\mtx{B}}(u)$, $s_{\mtx{C}}(u)$ and $s_{\Delta}(u)$ yielding a time-varying system. The authors also propose a hardware-aware implementation with linear complexity using an algorithm that takes advantage of GPU memory hierarchy. This novel time-variant version of SSMs, also called SSMs with Selective Scan or S6,  has the promise of improved context-awareness without losing the linear complexity and memory efficiency of time-invariant SSMs.

\subsection{SSMs in vision} 
\begin{figure}[t]
	\centering
	\includegraphics[width=0.9\linewidth]{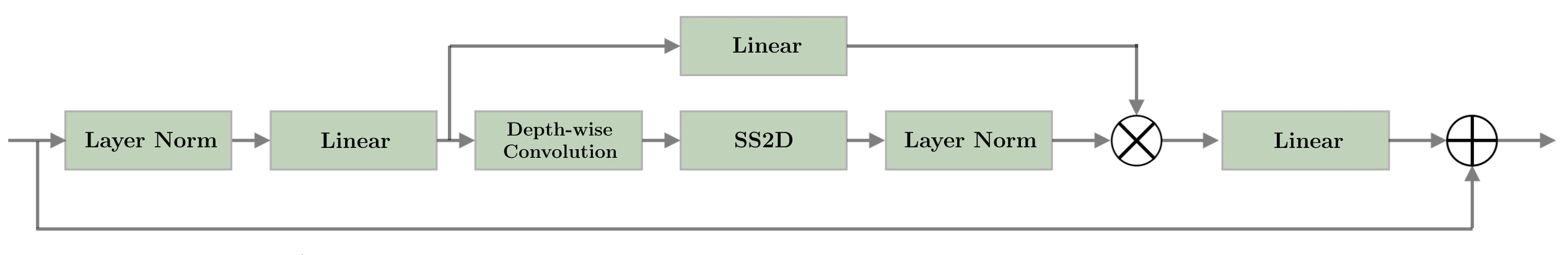}
	\caption{\label{fig:vss} The VSS block \cite{yu_vmamba_2024} scans the image along four different unrolled direction using state space models (SS2D). Linear layers are used to create feature embeddings, for gating and to produce the final output.}
\end{figure}

Motivated by the efficiency and success of SSMs in long-range tasks in NLP, there has been a recent flurry of activity in adapting state space models to vision, such as for image generation  \cite{yan_diffusion_2023} and for efficient vision backbones \cite{zhu_visionmamba_2024, yu_vmamba_2024}. The key challenge lies in the fact that, unlike in NLP and common $1$-D sequential tasks, images do not have an ordered direction and therefore it is unclear how to map them to a single sequence that SSMs can work with. Some works leverage bidirectional SSMs \cite{yan_diffusion_2023, wang2022pretraining}, where the image is flattened, processed in both directions via separate SSMs and the outputs combined to produce a final output sequence.
Authors in \cite{yu_vmamba_2024} propose flattening the image along four directions (top-left to bottom-right, top-right to bottom-left and both the opposite direction) and processing each sequence using individual S6 models, a technique they call a 2D selective scan (SS2D). Their fundamental computation block is the Visual State Space (VSS) block (Fig. \ref{fig:vss}), that processes linearly embedded features with a combination of SS2D and depth-wise convolution. 
	\section{Method}
In this work, we propose \archname{} for image restoration, an architecture with the efficiency of convolutional networks combined with the long-range modeling power of Transformers. We leverage S6 state space models with linear scaling in input size, allowing much faster training and inference compared to Transformer-based architectures, especially at high image resolutions where computing self-attention becomes infeasible.
\begin{figure}[t]
	\centering
	\includegraphics[width=0.9\linewidth]{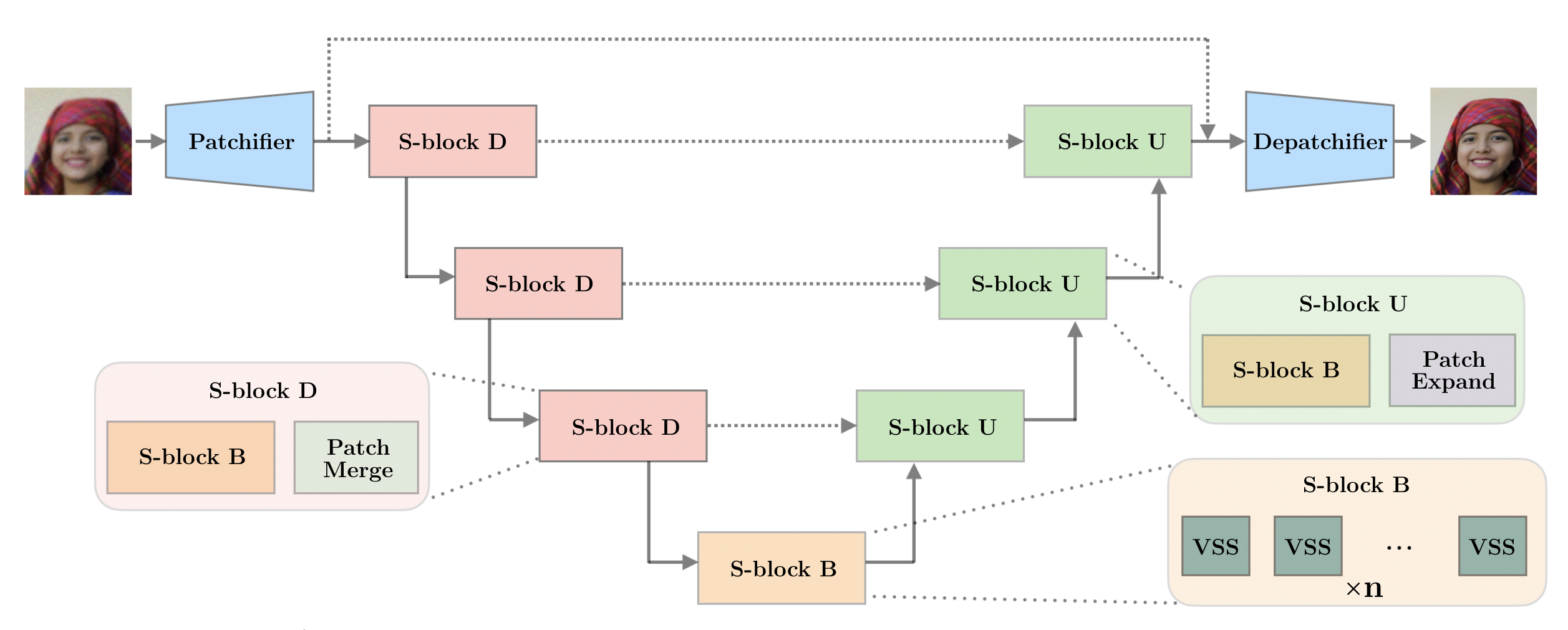}
	\caption{\label{fig:overview} Overview of our Serpent. Serpent has a U-Net architecture and use S-blocks at each layer. The S-block D is consist of $n$ VSS block in sequence.}
\end{figure}

\subsection{\archname{} architecture} An overview of \archname{} is depicted in Figure \ref{fig:overview}. First, the patchifier partitions the $H \times W \times C$ input image into patches of a particular width/height $P$, where each patch is subsequently embedded producing a representation with shape $H/P \times W/P \times D$. The embedding dimension $D$ is a hyperparameter that controls the overall capacity of the model. Inspired by the success of U-Nets \cite{ronneberger2015u} in low-level vision tasks in the general and medical domains, we design an architecture that processes the input signal on multiple scales using SSMs. Specifically, we apply a sequence of downsampling blocks that simultaneously merges image patches and increases their embedding dimension (or channels). The representation with the lowest spatial resolution forms a bottleneck in the network that encourages the model to extract a compact representation of the input image. A series of upsampling blocks then recovers the original input dimension from the compressed bottleneck representation. We introduce skip connections to aid with information flow between corresponding scales in the down- and upsampling branches.

\subsection{\archname{} block} The \archname{} block (\blockb{}) is our main processing block which consists of a stack of $n$ VSS blocks, each applying an S6 model with four-directional unrolling of the input image. With efficiency in mind, we allocate more VSS blocks at lower scales in order to reduce compute cost, as features are represented by shorter sequences at these scales. As the input sequence after unrolling the image consists of vectors of embedded patches, we apply separate SSMs along each embedding dimension without weight sharing. The downsampling block (\blockd{}) consists of \blockb{} followed by a patch merging operation, which is analogous to downsampling via strided convolutions in CNNs. Similarly, in the upsampling path we combine \blockb{} with a patch expanding operation forming \blocku{}.

\subsection{Merging and expanding patches} The patch merging block reduces the input along each spatial dimensions by a factor of $2$ and expands the number of channels by the same factor. First, we split the image into a grid of $2\times2$ patches and concatenate all pixels in a patch channel-wise. Then, we reduce the number of channels by a factor of $2$ with a linear layer.
The patch expanding block increases the input width and height by a factor of $2$ and decreases the number of channels by the same factor. To achieve this, we increase the channel dimension by $2$ using a linear layer. Then, we rearrange each pixel into $2\times2$ patches by dividing along the channel dimension, increasing the input width and height by a factor of $2$.

We define three main models: \textit{\archb}, \textit{\archl}, and \textit{\archh{}} with patch size $P=4$, $2$, and $1$ respectively, and fix $D$. Increasing the patch size makes the model faster, as the sequence length processed by SSMs is proportional to the number of image patches. However, as patches are mapped to the same channel dimension $D$, some information may be fundamentally lost with larger patch sizes. We also note that using the above scaling scheme, the model size in terms of number of parameters remains constant.
	\section{Experiments}
\subsection{Setup} 
 For \archname{} models, we set the embedding dimension in the patchifier to $D=32$. All models operate on $4$ scales: three consecutive up/down-sampling stages and one bottleneck stage. For each model, we have $[2, 2, 3, 3]$ VSS-blocks in \blockb{} of the first to the fourth layer. We set the hidden dimension in SSM blocks to $\frac{c}{6}$, where $c$ denotes the number of channels at the corresponding scale. We compare \archname{} to a fully-convolutional baseline, U-Net \cite{ronneberger2015u}, where we scale the model by varying the number of convolution filters (denoted as \textit{U-Net(channel number)}). Furthermore, we compare with a state-of-the-art image restoration method, SwinIR \cite{liang_swinir_2021}, that leverages a combination of SwinTransformer blocks and convolutions. We set the hidden dimension to $96$ for SwinIR-B and $180$ for SwinIR-L. Lastly, we compare with  Restormer \cite{zamir2022restormer}, a technique that leverages a channel-wise attention mechanism. We set the hidden dimension to 24 with 4 layers. From the first to the last layer, the number of Transformer blocks are $[4, 6, 6, 8]$, attention heads are $[1, 2, 4, 8]$, and number of channels are $[24, 48, 96, 192]$. We perform experiments on two image restoration tasks: (1) Gaussian deblurring on FFHQ \cite{karras_style-based_nodate} with kernel size of $61$ and additive Gaussian noise with $\sigma = 0.05$ and (2) $8\times$ super-resolution with the same amount of additive Gaussian noise as in the deblurring task. All models are trained for $60$ epochs with Adam, which has been sufficient for convergence. The learning rate used for training is $0.001$ for Serpent, $0.0005$ for U-Net, $0.003$  for Restormer, and $0.001$ for SwinIR.
\begin{figure}[t]
	\centering
	\includegraphics[width=0.9\linewidth]{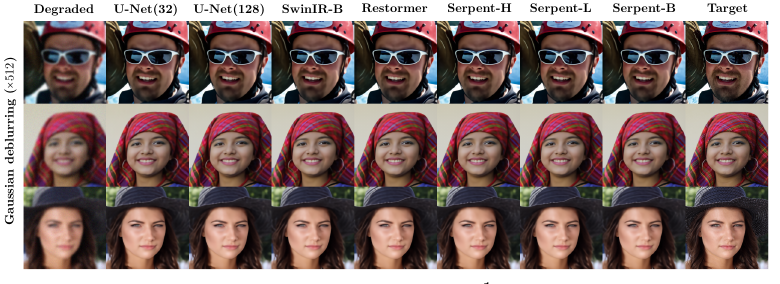}
	\includegraphics[width=0.9\linewidth]{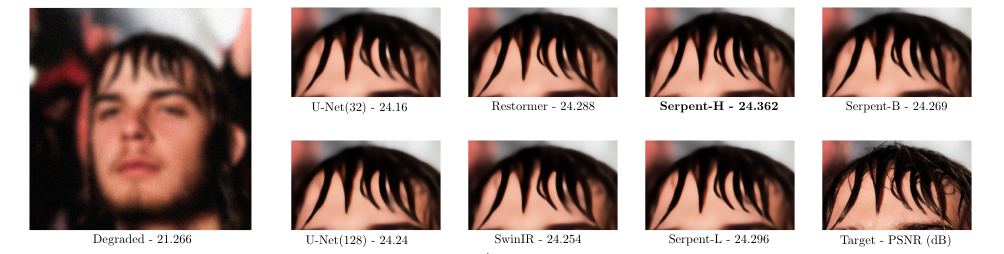}
	\caption{\label{fig:vis_compare}Visual comparison of reconstructions on the FFHQ $512\times$ deblurring task.}
\end{figure}

\begin{figure}[t]
	\centering
	\includegraphics[width=0.9\linewidth]{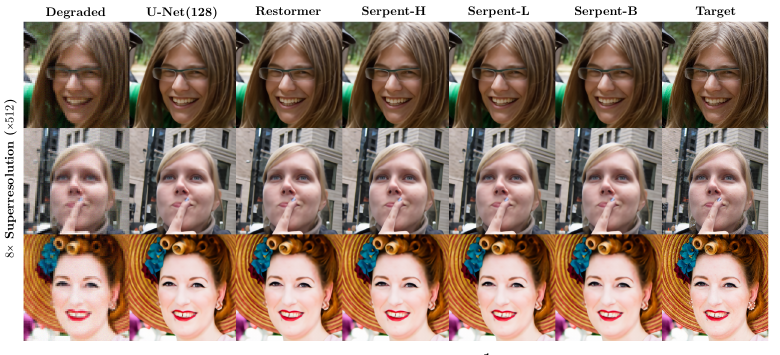}
	\caption{\label{fig:vis_compare_super}Visual comparison of reconstructions on the FFHQ $8\times$ superresolution task.}
\end{figure}

\textbf{Progressive learning --} Following the idea in \cite{zamir2022restormer} we employ progressive learning to improve the performance and reduce the training time of our models. We start training the model on smaller patches of the image (produced by random cropping) and gradually increase the resolution during training. By using this scheme, we can significantly reduce training time (at least by a factor of $2$) and improve the performance. We employ progressive learning for \archname{}, Restormer, and Unet. For SwinIR, since it only accepts fixed size images without further modifications, we cannot use progressive learning.

\textbf{Metrics --} We evaluate reconstruction performance via PSNR, SSIM and LPIPS. LPIPS is a perceptual metric that better reflects image quality as perceived by humans. We evaluate efficiency along four dimensions: compute cost in FLOPS, training time of a single epoch, number of model parameters and GPU memory cost during training. All the above metrics are evaluated on the same GPU (NVIDIA H100) with batch size $1$. 
\begin{table}[t]
	\centering
	\resizebox{12.5cm}{!}{
		\begin{tabular}{ lccccc }
			\toprule
			&\multicolumn{5}{c}{\textbf{Gaussian deblurring ($\times 256$)}}  \\
			\cmidrule{2-6}
			\textbf{Method} &PSNR$(\uparrow)$&SSIM$(\uparrow)$&LPIPS$(\downarrow)$ &GFLOPS & GPU mem. (GB)\\
			\midrule
			Restormer \cite{zamir2022restormer}&\underline{28.75}&\underline{0.8143}&\underline{0.2931}& 40.5 & 8.6\\
			SwinIR-L \cite{liang_swinir_2021}&28.74&0.8137&0.2938 & 746.9 & 23.5\\
			U-Net (32) \cite{ronneberger2015u}&28.32&0.8013&0.3040 & 12.2 & 2.6\\
			U-Net (128) \cite{ronneberger2015u}&28.51&0.8078&0.2955 & 192.8 & 4.7\\
			\midrule
			\archb{} (ours) &28.44&0.8054&0.3041 & 1.4 & 2.8\\
			\archl{} (ours)&28.64&0.8116&0.2950 & 5.6 & 3.4\\
			\archh{} (ours)&\textbf{28.77}&\textbf{0.8149}&\textbf{0.2911} & 22.2 & 5.3\\
			\bottomrule
		\end{tabular}
	}
	\resizebox{12.5cm}{!}{
		\begin{tabular}{ lccccc }
			\toprule
			&\multicolumn{5}{c}{\textbf{Gaussian deblurring ($\times 512$)}} \\
			\cmidrule{2-6}
			\textbf{Method} &PSNR$(\uparrow)$&SSIM$(\uparrow)$&LPIPS$(\downarrow)$ &GFLOPS & GPU mem. (GB)\\
			\midrule
			Restormer \cite{zamir2022restormer}&\textbf{28.51}&\underline{0.7797}&\underline{0.4136} & 161.9 & 28.5\\
			SwinIR-B \cite{liang_swinir_2021}&{28.37}&0.7756&0.4214 & 889.5 & 51.8\\
			U-Net (32) \cite{ronneberger2015u}&28.23 &0.7710&0.4162 & 48.6 & 3.7\\
			U-Net (128) \cite{ronneberger2015u}&28.35 &0.7751&0.4138 & 771.1 & 7.8\\
			\midrule
			\archb{} (ours)&28.35 &{0.7755}&{0.4195} & 5.6 & 4.1\\
			\archl{}  (ours)&\underline{28.48} &0.7790&\underline{0.4136}& 22.2 & 6.4\\
			\archh{}  (ours)&\textbf{28.51} &\textbf{0.7800}&\textbf{0.4127} & 88.8 & 15.9\\
			\bottomrule
		\end{tabular}
	}
	\caption{Reconstruction performance comparison on FFHQ deblurring. \underline{Top:} at $256\times$ resolution \archname{} outperforms state-of-the-art Restormer, SwinIR, and convolutional baselines. \underline{Bottom:} at higher resolution ($512\times$), our proposed method surpasses SwinIR in every metric with orders of magnitude lower compute and memory cost.\label{tab:perf}}
\end{table}

\begin{table}[t]
	\centering
	\resizebox{8cm}{!}{
		\begin{tabular}{ lccc }
			\toprule
			&\multicolumn{3}{c}{\textbf{$\times 8$ Superresolution ($\times 512$)}} \\
			\cmidrule{2-4}
			\textbf{Method} &PSNR$(\uparrow)$&SSIM$(\uparrow)$&LPIPS$(\downarrow)$ \\
			\midrule
			Restormer \cite{zamir2022restormer}&\underline{29.20}&\underline{0.7984}&\underline{0.3809}\\
			U-Net (128) \cite{ronneberger2015u}&28.95 &0.7925&0.3827\\
			\midrule
			\archb{} (ours)&29.03 &{0.7940}&{0.3871}\\
			\archl{}  (ours)&29.16 &0.7973&0.3815\\
			\archh{}  (ours)&\textbf{29.23} &\textbf{0.7988}&\textbf{0.3800}\\
			\bottomrule
		\end{tabular}
	}
	\caption{Reconstruction performance comparison on FFHQ Superresolution. Similar to deblurring tasks, our proposed method out-performs state-of-the-art Restormer while having similar computation cost and lower memory cost.\label{tab:sref}}
\end{table}

 \subsection{Performance results} 
 Our results on reconstruction performance are summarized in Table \ref{tab:perf}.  At $256\times$ resolution, \textit{\archb{}} has comparable or lower compute and memory requirements as lightweight convolutional baselines, but with improved reconstruction performance.  \textit{\archl{}} is a higher capacity model that significantly outperforms the largest convolutional baseline, but with highly reduced FLOPS. Finally,  \textit{\archh{}} is our best performing model that surpasses SwinIR in every metric (for compute and performance). Its performace and compute are on par with Restormer, but with greatly reduced compute and memory requirements. The advantage introduced by efficient modeling of long-range dependencies becomes even more dominant at $512\times$ resolution. \archl{} outperforms SwinIR and convolutional baselines in every image quality metric and matches the performance of Restormer, but with orders of magnitude less compute and memory. Reconstructed samples are depicted in Fig. \ref{fig:vis_compare}.
 
 \begin{figure}
 	\centering
 	\begin{subfigure}{0.42\textwidth}
 		\centering
 		\includegraphics[width=\textwidth]{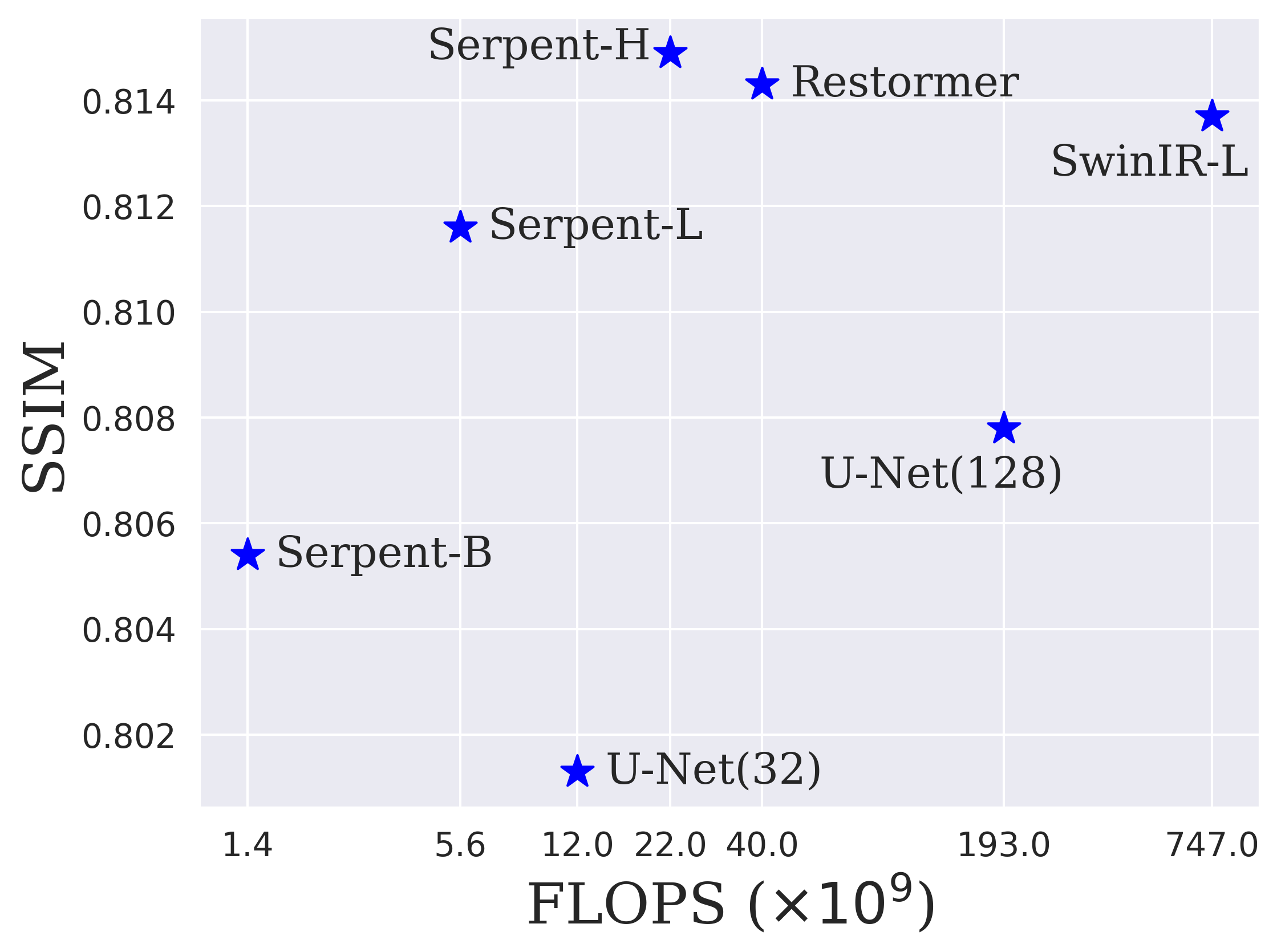}
 		\caption{Compute comparison \label{fig:eff256_comp}}
 	\end{subfigure}
 	\begin{subfigure}{0.42\textwidth}
 		\centering
 		\includegraphics[width=\textwidth]{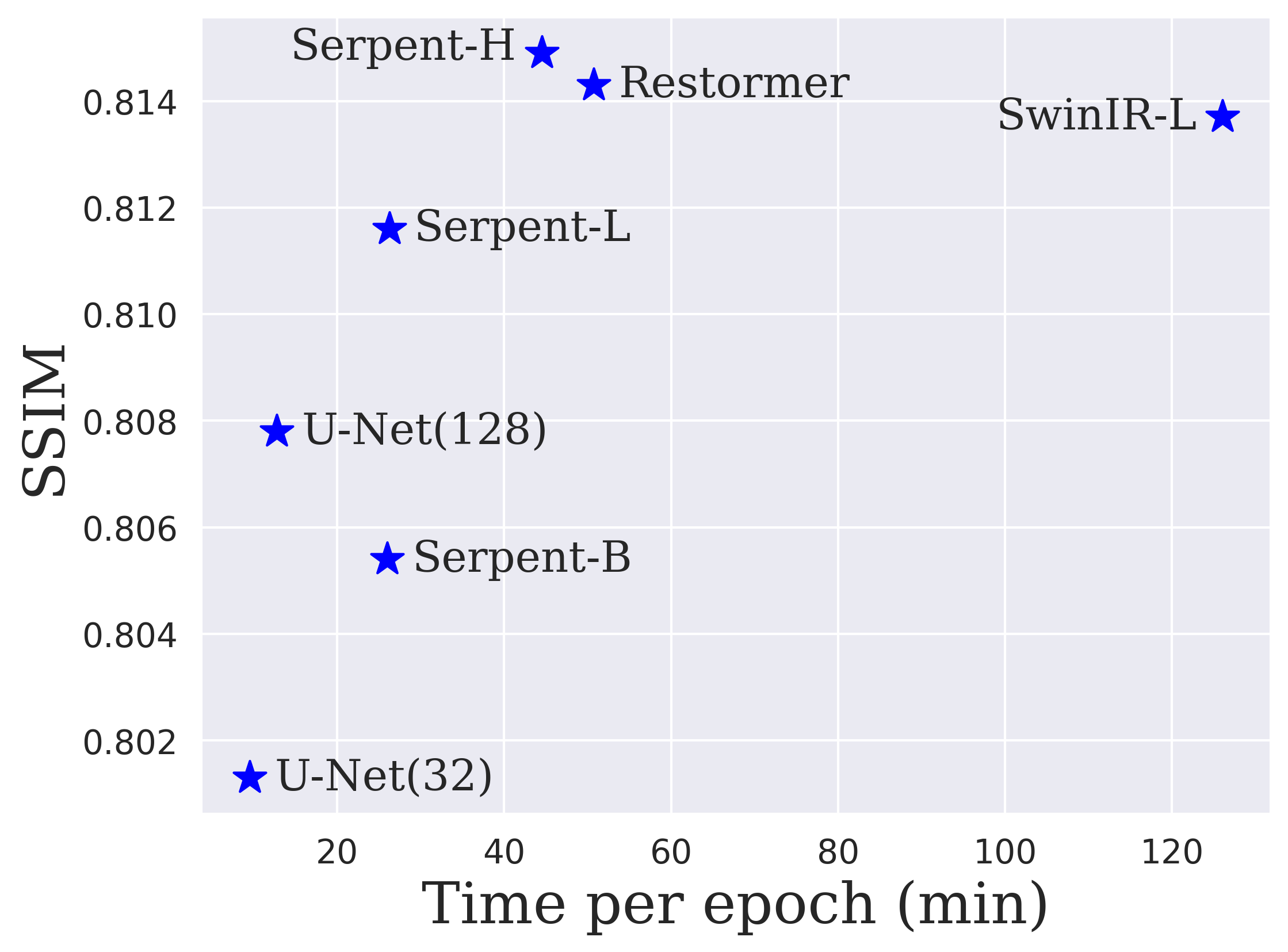}
 		\caption{Timing comparison\label{fig:eff256_time}}
 	\end{subfigure}
 	\\
 	\begin{subfigure}{0.42\textwidth}
 		\centering
 		\includegraphics[width=\textwidth]{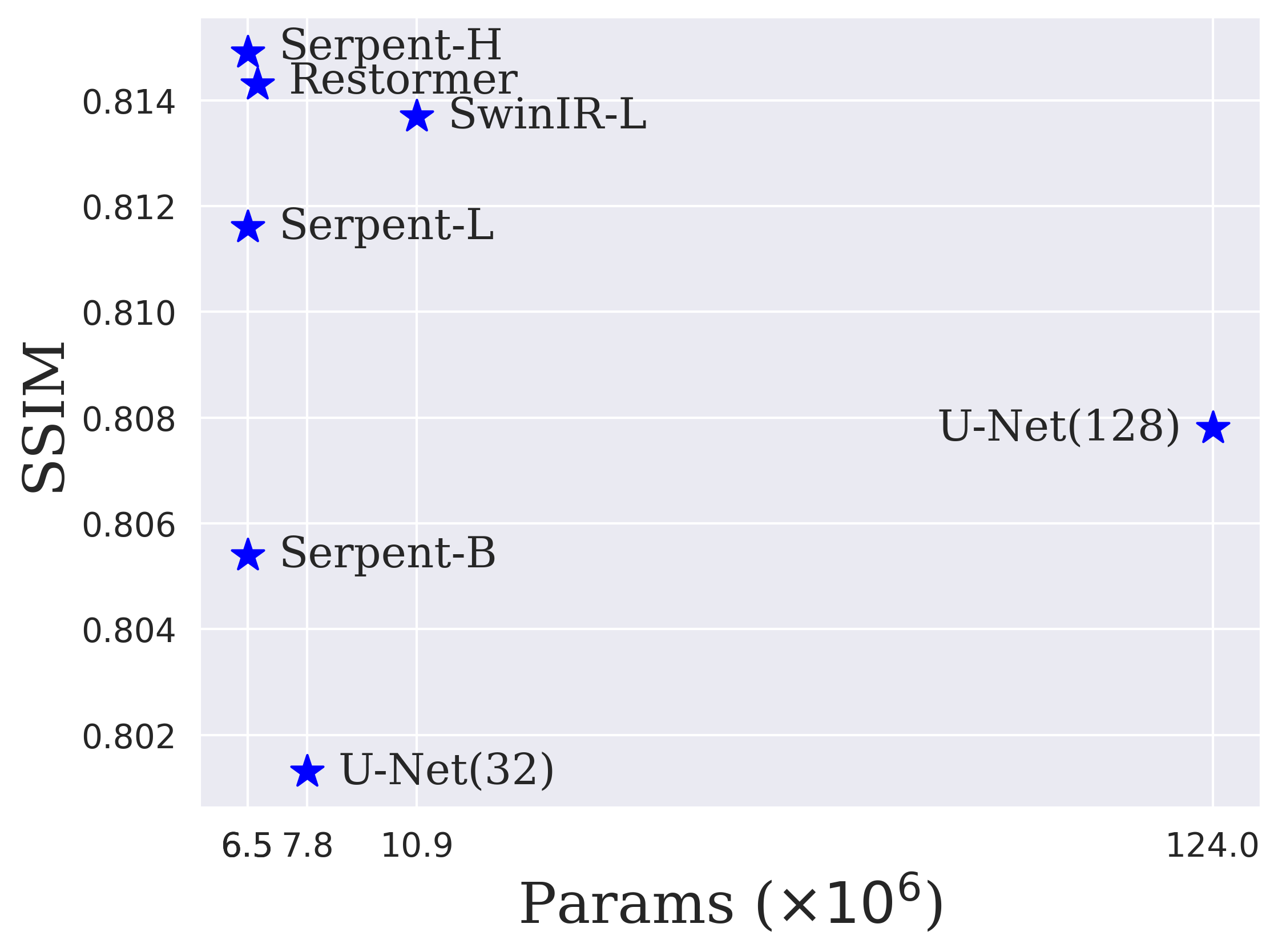}
 		\caption{Model size comparison\label{fig:eff256_params}}
 	\end{subfigure}
 	\begin{subfigure}{0.42\textwidth}
 		\centering
 		\includegraphics[width=\textwidth]{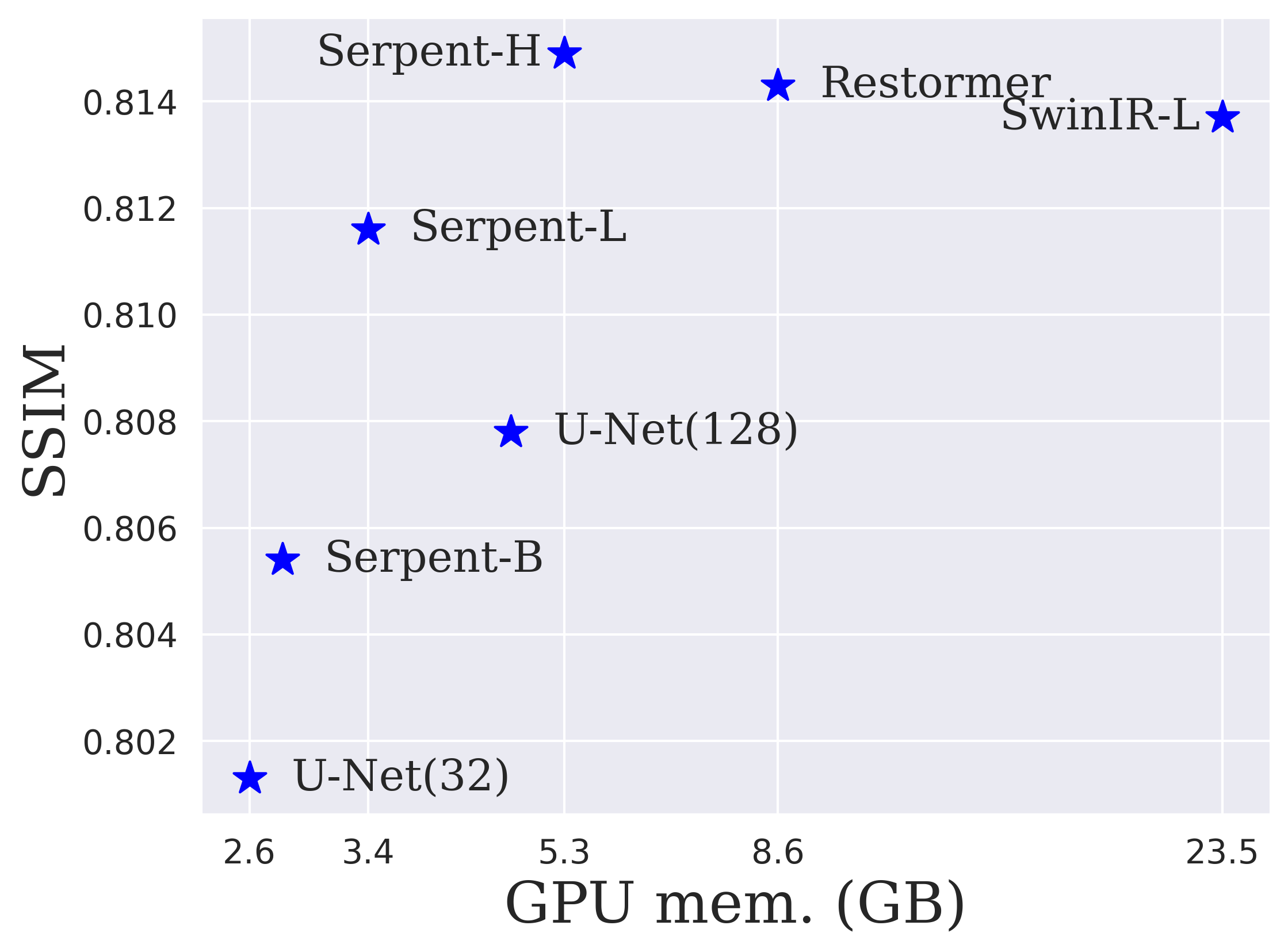}
 		\caption{Memory comparison\label{fig:eff256_mem}}
 	\end{subfigure}
 	\begin{subfigure}{0.42\textwidth}
 		\centering
 		\includegraphics[width=\textwidth]{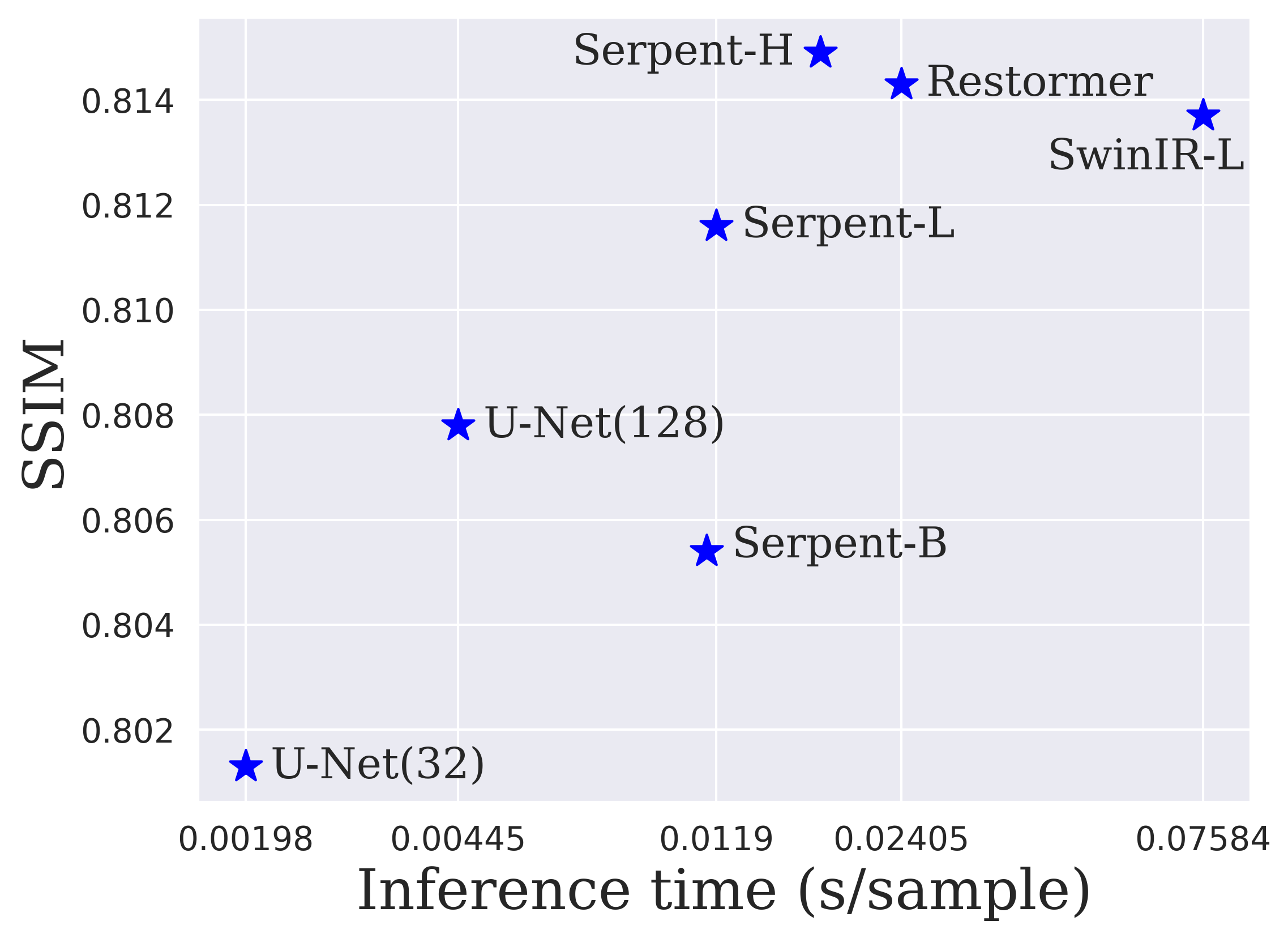}
 		\caption{Inference comparison\label{fig:eff256_inf}}
 	\end{subfigure}
 	\caption{Efficiency of \archname{}: comparison across various metrics (compute,  training time, model size, memory, inference time) with popular methods on FFHQ ($\times 256$) deblurring. \label{fig:eff256}\vspace{-0.4cm}}
 \end{figure}
 
 \begin{figure}
 	\centering
 	\begin{subfigure}{0.42\textwidth}
 		\centering
 		\includegraphics[width=\textwidth]{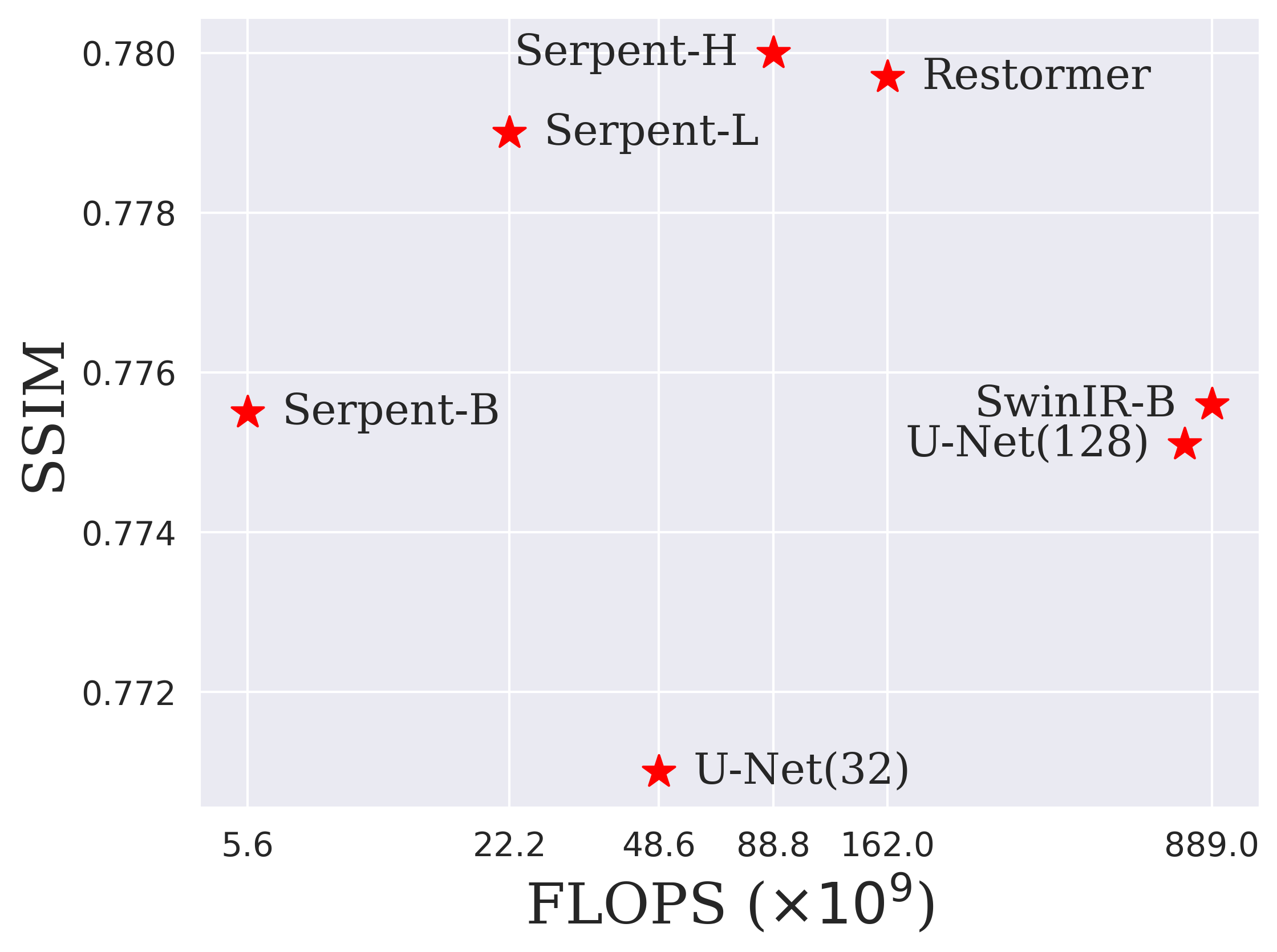}
 		\caption{Compute comparison  \label{fig:eff512_comp}}
 	\end{subfigure}
 	\begin{subfigure}{0.42\textwidth}
 		\centering
 		\includegraphics[width=\textwidth]{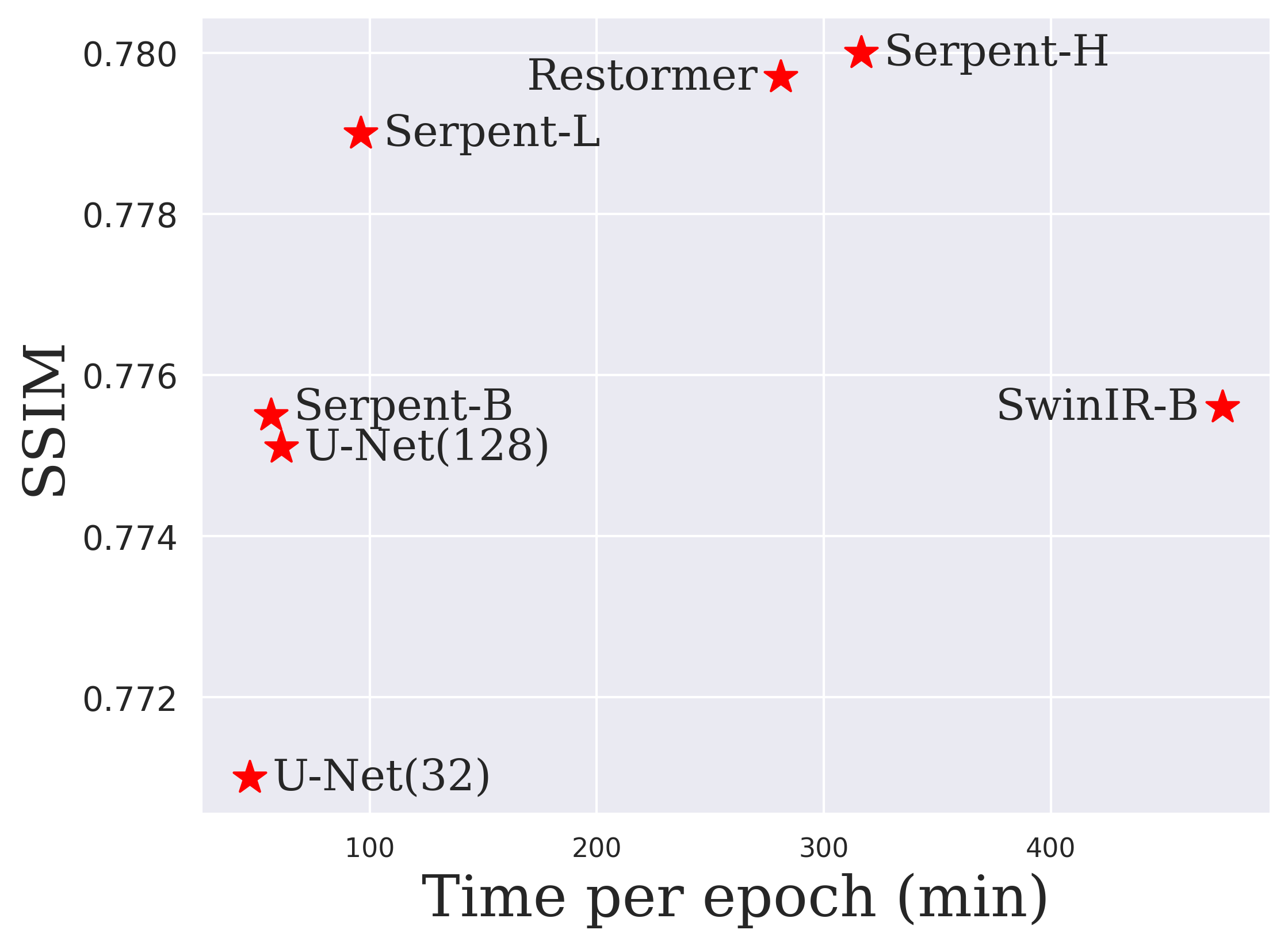}
 		\caption{Timing comparison\label{fig:eff512_time}}
 	\end{subfigure}
 	\\
 	 	\begin{subfigure}{0.42\textwidth}
 		\centering
 		\includegraphics[width=\textwidth]{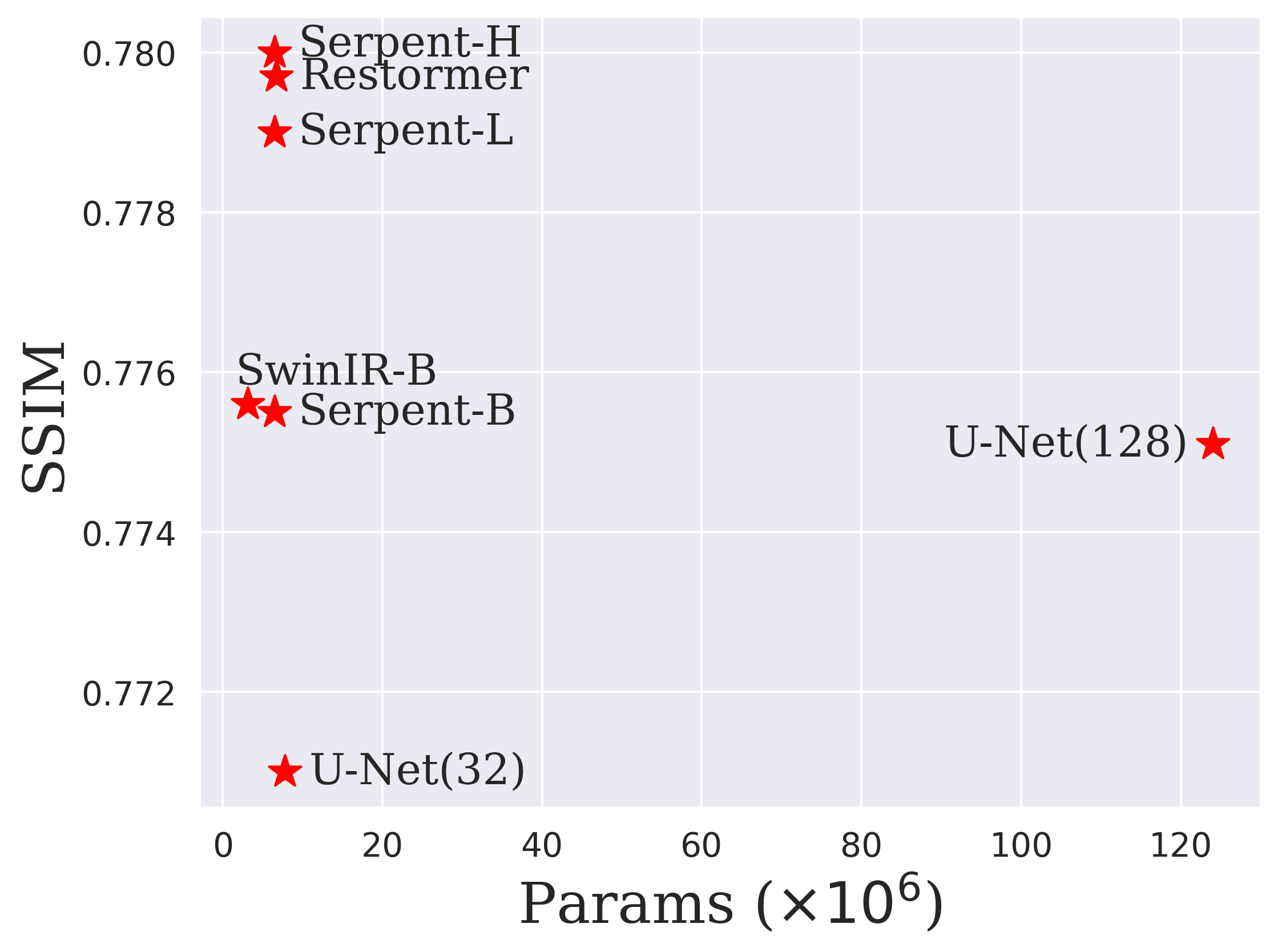}
 		\caption{Model size comparison\label{fig:eff512_params}}
 	\end{subfigure}
 	\begin{subfigure}{0.42\textwidth}
 		\centering
 		\includegraphics[width=\textwidth]{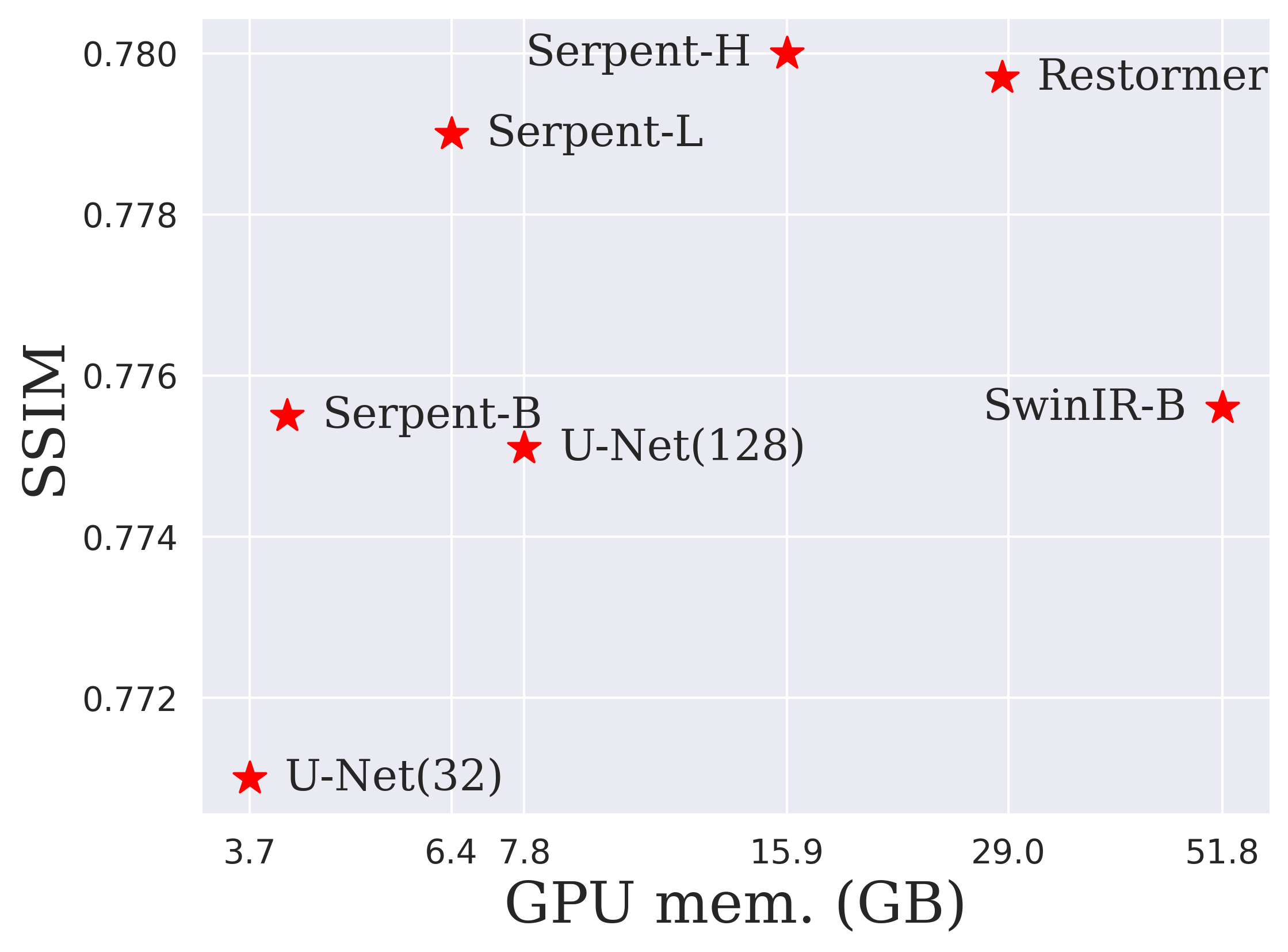}
 		\caption{Memory comparison  \label{fig:eff512_mem}}
 	\end{subfigure}
 	\begin{subfigure}{0.42\textwidth}
 		\centering
 		\includegraphics[width=\textwidth]{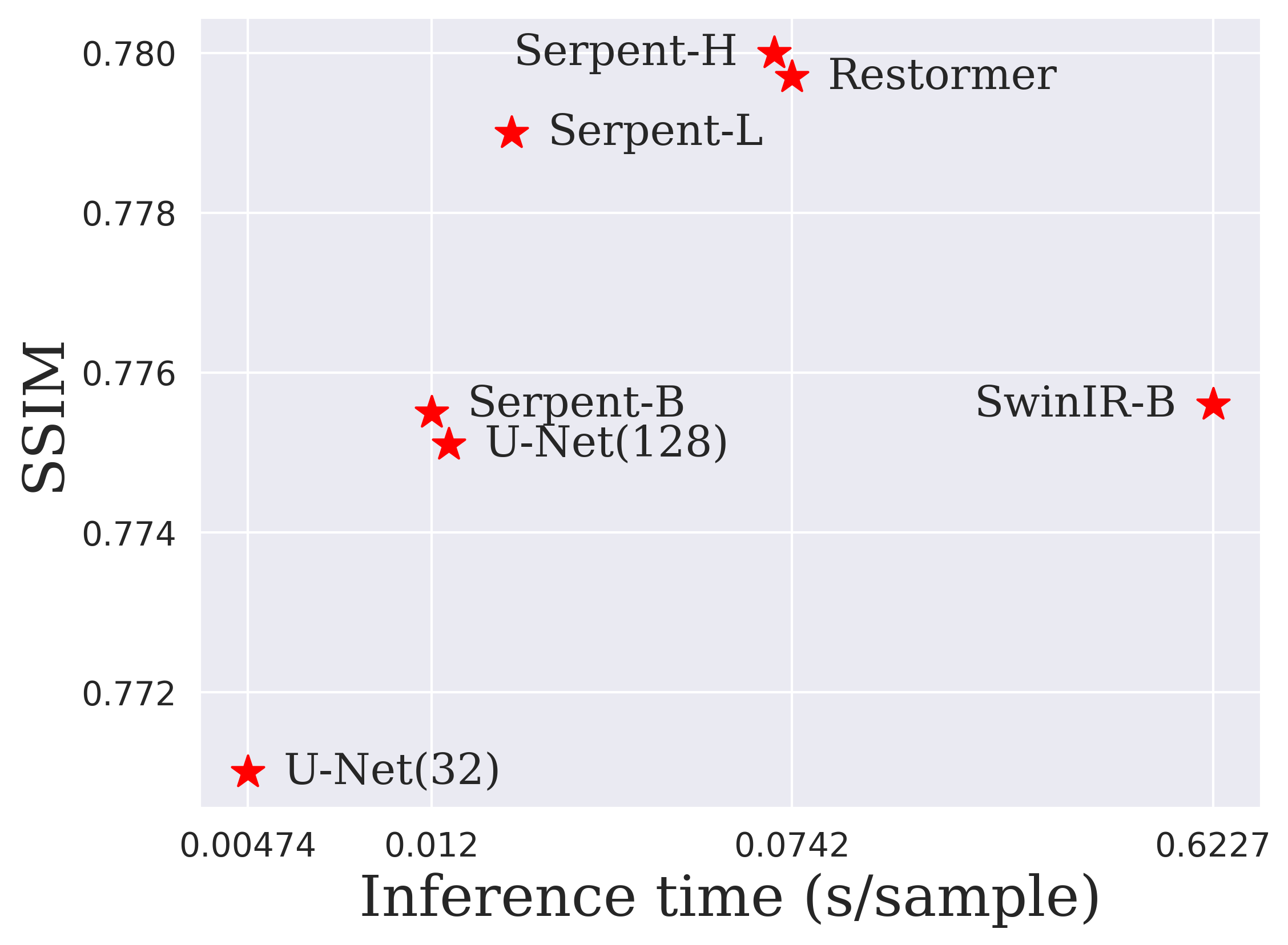}
 		\caption{Inference comparison  \label{fig:eff512_inf}}
 	\end{subfigure}
 	\caption{Efficiency of \archname{} at high resolution: comparison with popular methods on FFHQ ($\times 512$) deblurring. \label{fig:eff512}\vspace{-0.4cm} }
 \end{figure}
 
 \subsection{Efficiency results}
 \textbf{Compute efficiency --} \underline{At $256\times$ resolution}, \archh{} surpasses SwinIR-L performance while only requiring $3\%$ of the inference cost of SwinIR-L in terms of FLOPS (Fig. \ref{fig:eff256_comp}). Moreover, training time is greatly reduced: \archh{} training takes close to $30\%$ of SwinIR-L training per epoch (Fig. \ref{fig:eff256_time}). Compared to U-Net baselines, \archl{} greatly outperforms even the largest convolutional baseline with only $3\%$ of FLOPS compared to U-Net(128). Furthermore, \archh{} matches Restormer performance with about $50\%$ of FLOPS. On the other hand, we observe that per epoch training time of U-Net baselines  is lower than that of \archname{} models, which we hypothesize is due to the more efficient low-level GPU implementation of convolutional layers. \underline{At $512\times$ resolution}, \archl{} outperforms SwinIR-B while reducing FLOPS by a factor of $40$ (Fig. \ref{fig:eff512_comp}).  Strikingly, \archb{} achieves comparable performance to SwinIR-B with a factor of $150$ reduction in FLOPS. Similarly, \archl{} has a close performance to Restormer, but with $7$ times less FLOPS and $3$ times less training time. Moreover, at this resolution, \archb{} has performance on par with the larger U-Net, but with lower compute cost. Compute scaling in terms of FLOPS with respect to input resolution is summarized in Figure \ref{fig:frontpage} (left).
 
 \textbf{Parameter efficiency--} In terms of model size, \archname{} models have less than half of the parameters of SwinIR-L and approximately $60\%$ of the parameters as the smallest U-Net baseline. \archname{} models and Restormer have a comparable number of parameters (Fig. \ref{fig:eff256_params}). Despite the small size, \archh{} is able to match the performance of SwinIR-L and outperform all U-Net baselines.
 
 \textbf{Memory efficiency--} GPU memory requirements constitute a key bottleneck when scaling image reconstruction techniques to higher resolutions. For instance, at $512\times$ resolution, we are unable to train SwinIR-L with batch size 1 and without activation checkpointing on the most advanced H100 GPUs with $80$ GB memory. Overall, we observe that our largest models require $4-6\times$ less GPU memory to train compared to SwinIR (Figures \ref{fig:eff256_mem} and \ref{fig:eff512_mem}), have comparable memory requirements to fully-convolutional U-Nets with efficient and optimized low-level GPU implementation. In addition, \archh{} memory requirement about $30\%$ less compared to Restormer. The favorable scaling in terms of GPU memory utilization with respect to input dimension is summarized in Figure \ref{fig:frontpage} (right).
	\section{Conclusion}
In this paper, we introduce \archname{}, an efficient image restoration network architecture utilizing structured state space models in a multi-scale fashion. Our experiments demonstrate the promising characteristics of \archname{} in terms of compute and memory efficiency, as well as model size. In particular, we observe up to a factor of $150\times$ reduction in FLOPS compared to state-of-the-art techniques on high-resolution images, with GPU memory requirements reduced by approximately a factor of $5\times$. A current limitation of \archname{} is a less substantial low-level software and hardware support for selective scans when compared with traditional convolution and attention operations. We believe that improved GPU-level support of SSMs would allow even further gains in efficiency over attention-based architectures.  Finally, we point out that \archname{} is a supervised machine learning technique, therefore the reconstructed images may show biases present in the training dataset. In order to avoid potential negative societal impact, it is crucial to train \archname{} on datasets curated with fairness considerations.
	{
		\small
	\bibliographystyle{ieee_fullname}
		\bibliography{references}
	}
	\newpage
\appendix
\section{Further training details}
\textbf{Training --} For the training, we used FFHQ \cite{karras_style-based_nodate} which has $60,000$ training and $10,000$ validation samples. We observed that $60$ epochs is enough for all models to converge in training. For the learning rate, we chose the best learning rate among 0.001, 0.0005, 0.0003, and 0.0001 for each model. For progressive learning, we use three different image size, at each step we double the resolution, and we train on the full-resolution training samples at the last step. 

\textbf{Metrics --} We use VMamba \cite{yu_vmamba_2024} source code\footnote{\url{https://github.com/MzeroMiko/VMamba}}, and \texttt{fvcore} to calculate FLOPS. For the other metrics, we train the models for 2 epochs on one H100 GPU without predicting validation samples, and measure memory consumption and training time.

\textbf{Training time --} We trained our models with H100 GPUs. For $\times 256$ images, it takes $95$ GPU hours to train \archh, and for $\times 512$ images, it needs $179$ GPU hours to train.
%
\end{document}